\documentclass{article}
\usepackage[]{graphicx}\usepackage[]{color}
\makeatletter
\def\maxwidth{ %
  \ifdim\Gin@nat@width>\linewidth
    \linewidth
  \else
    \Gin@nat@width
  \fi
}
\makeatother

\definecolor{fgcolor}{rgb}{0.345, 0.345, 0.345}

\usepackage{framed}
\makeatletter
 {\par\unskip\endMakeFramed%
 \at@end@of@kframe}
\makeatother

\definecolor{shadecolor}{rgb}{.97, .97, .97}
\definecolor{messagecolor}{rgb}{0, 0, 0}
\definecolor{warningcolor}{rgb}{1, 0, 1}
\definecolor{errorcolor}{rgb}{1, 0, 0}
\newenvironment{knitrout}{}{} 

\usepackage{alltt} 
\usepackage[american]{babel}
\usepackage[utf8]{inputenc}
\usepackage{csquotes}

\usepackage{setspace}

\usepackage{amsmath,amssymb,amsfonts}

\usepackage{url}   
\usepackage{graphicx}   
\usepackage{verbatim}   
\usepackage{caption}
\usepackage{pdflscape}

\usepackage{fancyvrb}

\usepackage{newfloat}
\DeclareFloatingEnvironment[
]{listing}

\title{Bayesian linear mixed models using Stan:
A tutorial for psychologists, linguists, and cognitive scientists}

\author{Tanner Sorensen$^{1}$ and Shravan Vasishth$^{1,2}$\\
1: Department of Linguistics, University of Potsdam, Germany\\
2: School of Mathematics and Statistics, Sheffield University, UK\\
Correspondence to: \{tanner.sorensen,vasishth\}@uni-potsdam.de}

\IfFileExists{upquote.sty}{\usepackage{upquote}}{}
\begin{document}

\maketitle

\begin{abstract}
With the arrival of the R packages \texttt{nlme} and \texttt{lme4}, linear mixed models (LMMs) have come to be widely used in experimentally-driven areas like psychology, linguistics, and cognitive science. This tutorial provides a practical introduction to fitting LMMs in a Bayesian framework using the probabilistic programming language Stan. We choose Stan (rather than WinBUGS or JAGS) because it provides an elegant and scalable framework for fitting models in most of the standard applications of LMMs. We ease the reader into fitting increasingly complex LMMs, first using a two-condition repeated measures self-paced reading study, followed by a more complex $2\times 2$ repeated measures factorial design that can be generalized to much more complex designs.
\end{abstract}

\section{Introduction}

Linear mixed models, or hierarchical/multilevel linear models, have become the main workhorse of experimental research in psychology, linguistics, and cognitive science, where repeated measures designs are the norm.
Within the programming environment R~\cite{R}, 
the \texttt{nlme} package \cite{pinheirobates} and its successor, \texttt{lme4} \cite{batesetal2014b} have revolutionized the use of linear mixed models (LMMs) due to their simplicity and speed: one can fit fairly complicated models relatively quickly, often with a single line of code. A great advantage of LMMs over traditional approaches such as repeated measures ANOVA and paired t-tests is that there is no need to aggregate over subjects and items to compute two sets of F-scores (or several t-scores) separately; a single model can take all sources of variance into account simultaneously. Furthermore, comparisons between conditions can easily be implemented in a single model through appropriate contrast coding.

Other important developments related to LMMs have been unfolding in  computational statistics.
Specifically, probabilistic programming languages like WinBUGS \cite{lunn2000winbugs}, JAGS \cite{plummer2011jags} and Stan \cite{stan-manual:2014}, among others, have  made it possible to fit Bayesian LMMs quite easily. However, one prerequisite for using these programming languages is that some background statistical knowledge is needed before one can define the model. This difficulty is well-known; for example, Spiegelhalter and colleagues \cite[4]{spiegelhalter2004bayesian} write: ``Bayesian statistics has a (largely deserved) reputation for being mathematically challenging and difficult to put into practice\dots''. 

The purpose of this paper is to facilitate a first encounter with model specification in one of these programming languages, Stan. The tutorial is aimed primarily at psychologists, linguists, and cognitive scientists who have used \texttt{lme4} to fit models to their data, but may have only a basic knowledge of the underlying LMM machinery. A diagnostic test is that they may not be able to answer some or all of these questions: what is a design matrix; what is contrast coding; what is a random effects variance-covariance matrix in a linear mixed model? Our tutorial is not intended for statisticians or psychology researchers who could, for example, write their own Markov Chain Monte Carlo samplers in R or C++ or the like; for them, the Stan manual is the optimal starting point. The present tutorial attempts to ease the beginner into their first steps towards fitting Bayesian linear mixed models. More detailed presentations about linear mixed models are available in several textbooks; references are provided at the end of this tutorial. 

We have chosen Stan as the programming language of choice (over JAGS and WinBUGS) because it is possible to fit arbitrarily complex models with Stan. For example, it is possible (if time consuming) to fit a model with $14$ fixed effects predictors and two crossed random effects by subject and item, each involving a $14\times 14$ variance-covariance matrix \cite{BatesEtAlParsimonious}; as far as we are aware,  such models cannot, as far as we know, be fit in JAGS or WinBUGS.\footnote{Whether it makes sense in general to fit such a complex model is a different issue; see \cite{Gelman14}, and \cite{BatesEtAlParsimonious} for recent discussion.}

In this tutorial, 
we take it as a given that the reader is interested in learning how to fit Bayesian linear mixed models. We do not try to explain the advantages this approach affords beyond the classical frequentist approach; for such justification, the reader is directed to the rich literature relating to a comparison between Bayesian versus frequentist statistics (such as the provocatively titled paper by Lavine \cite{lavine1999bayesian}; and the highly accessible textbook by 
Kruschke \cite{kruschke2014doing}). 
We also assume that the reader is aware of Bayes' Theorem, which for our purposes amounts to the statement that the posterior distribution is proportional to the prior times the likelihood.  For the purposes of this paper, the goal of a Bayesian analysis is simply to derive the posterior distribution of each parameter of interest, given some data and prior beliefs about the distributions of the parameters.

The tutorial is structured as follows. We begin by 
successively building up increasingly complex LMMs using the data-set reported by~Gibson and Wu\cite{gibsonwu}, which has a  simple two-condition design. At each step, we explain the structure of the model.  The next section takes up inference for this two-condition design.  Then we demonstrate how one can fit a somewhat more complex $2\times 2$ factorial design.

This paper was written using a literate programming tool, \texttt{knitr} \cite{xie2013knitr}; this integrates documentation for the accompanying code with the paper.
The \texttt{knitr} file that generated this paper, as well as all the code and data used in this tutorial, can be downloaded from our website: 

http://www.ling.uni-potsdam.de/$\sim$vasishth/statistics/BayesLMMs.html

We start with the two-condition repeated measures data-set~\cite{gibsonwu} as a concrete running example. This simple example serves as a starter kit for fitting commonly used LMMs in the Bayesian setting. We assume that the reader has the relevant software installed; specifically, \texttt{rstan} in R. For detailed instructions, see  

https://github.com/stan-dev/rstan/wiki/RStan-Getting-Started

\section{Example 1: A two-condition repeated measures design}
\label{sec:modeling}

This section motivates the LMM with the self-paced reading data-set of~Gibson and Wu \cite{gibsonwu}. We introduce the data-set, state our modeling goals here, and proceed to build up increasingly complex LMMs.

\paragraph{The scientific question}
Subject and object relative clauses have been widely used in reading studies to investigate sentence comprehension processes. A subject relative is a sentence like \textit{The senator who interrogated the journalist resigned} where a noun (\textit{senator}) is modified by a relative clause (\textit{who interrogated the journalist}), and the modified noun is the grammatical subject of the relative clause. In an object relative, the noun modified by the relative clause is the grammatical object of the relative clause (e.g., \textit{The senator who the journalist interrogated resigned}). In both cases, the noun that is modified (\textit{senator}) is called the head noun.

A typical finding for English is that subject relatives are easier to process than object relatives~\cite{just1992ctc}. Natural languages generally have relative clauses, and the subject relative advantage has until recently been considered to be true cross-linguistically. However, Chinese relative clauses apparently represent an interesting counter-example to this generalization; recent work by~Hsiao and Gibson \cite{hsiao03} has suggested that in Chinese, \textit{object} relatives are easier to process than subject relatives at a particular point in the sentence (the head noun of the relative clause). We now present an analysis of a subsequently published data-set~\cite{gibsonwu} that evaluates this claim. 

\paragraph{The data}
The dependent variable of the experiment of~Gibson and Wu \cite{gibsonwu} was the reading time $\hbox{\texttt{rt}}$ of the head noun of the relative clause. This was recorded in two conditions (subject relative and object relative), with $37$ subjects and $15$ items, presented in a standard Latin square design. There were originally $16$ items, but one item was removed, resulting in $37\times 15=555$ data points. However, eight data points from one subject (id 27) were missing. As a consequence, we have a total of $555-8=547$ data points. The first few lines from the data frame are shown in Table~\ref{tab:dataframe1}; ``o'' refers to object relative and ``s'' to subject relative.

\begin{table}[ht]
\centering
\begin{tabular}{rrrlr}
  \hline
row & subj & item & so & rt \\ 
  \hline
1 & 1 &  13 & o & 1561 \\ 
2 &  1 &   6 & s & 959 \\ 
3 &  1 &   5 & o & 582 \\ 
4 &  1 &   9 & o & 294 \\ 
5 &  1 &  14 & s & 438 \\ 
6 &  1 &   4 & s & 286 \\ 
   \vdots & \vdots & \vdots & \vdots \\ 
547 & 9 & 11 & o & 350 \\   
   \hline
\end{tabular}
\caption{First six rows, and the last row, of the data-set of Gibson and Wu (2013), as they appear in the data frame.}\label{tab:dataframe1}
\end{table}

We build up the Bayesian LMM from a fixed effects model 
to a varying intercepts model 
and finally to a varying intercepts, varying slopes model 
(the ``maximal model'' of Barr and colleagues \cite{barr2011random}). The result is a probability model that expresses how the dependent variable, the reading time labeled $\hbox{\texttt{rt}}$, was generated in the experiment of~Gibson and Wu \cite{gibsonwu}. 

As mentioned above, 
the goal of Bayesian modeling is to derive the \textit{posterior probability distribution} of the model parameters from a \textit{prior probability distribution} and a \textit{likelihood function}. Stan makes it easy to compute this posterior distribution of each parameter of interest. 
The posterior distribution reflects what we should believe, given the data, regarding the value of that parameter.

\subsection{Fixed Effects Model (Simple Linear Model)}
\label{subsec:fixef}

We begin by making the working assumption that the dependent variable of reading time $\hbox{\texttt{rt}}$ on the head noun is approximately log-normally distributed~\cite{rouder2005}. This assumes that the logarithm of $\hbox{\texttt{rt}}$ is approximately normally distributed. The logarithm of the reading times, $\log \hbox{\texttt{rt}}$, has some unknown grand mean $\beta _0$. 
The mean of the log-normal distribution of $\hbox{\texttt{rt}}$ is the sum of $\beta _0$ and an adjustment $\beta _1\times \hbox{\texttt{so}}$ whose magnitude depends on the categorical predictor $\hbox{\texttt{so}}$, which has the value $-1$ when $\hbox{\texttt{rt}}$ is from the subject relative condition, and $+1$ when $\hbox{\texttt{rt}}$ is from the object relative condition. One way to write the model in terms of the logarithm of the reading times is as follows:

\begin{equation}\label{eq:fixef}
\log \hbox{\texttt{rt}}_{i} = \beta _0 + \beta _1\hbox{\texttt{so}}_i + \varepsilon_{i} 
\end{equation}

The index $i$ represents the $i$-th row in the data-frame (in this case, $i=1,\dots,547$); the term $\varepsilon_i$ represents the error in the i-th row.
With the above $\pm 1$ contrast coding, $\beta _0$ represents the grand mean of $\log \hbox{\texttt{rt}}$, regardless of relative clause type. It can be estimated by simply taking the grand mean of $\log \hbox{\texttt{rt}}$.
The parameter $\beta _1$ is an adjustment to $\beta _0$ so that the mean of $\log \hbox{\texttt{rt}}$ is $\beta _0 + 1 \beta _1$ when $\log \hbox{\texttt{rt}}$ is from the object relative condition, and $\beta _0 - 1 \beta _1$ when $\log \hbox{\texttt{rt}}$ is from the subject relative condition. Notice that $2\times \beta_1$ will be the difference in the means between the object and subject relative clause conditions.
Together, $\beta _0$ and $\beta _1$ make up the part of the model which characterizes the effect of the experimental manipulation, relative clause type (\texttt{so}), on the dependent variable \texttt{rt}. We call this a fixed effects model because we estimate the $\beta$ parameters, which are unvarying from subject to subject and from item to item. In R, this would correspond to fitting a simple linear model using the \texttt{lm} function, with \texttt{so} as predictor and $\log rt$ as dependent variable.

The error $\varepsilon $ is positive when $\log \hbox{\texttt{rt}}_i$ is greater than the expected value $\mu_i = \beta _0 + \beta _1 \hbox{\texttt{so}}_i$ and negative when $\log \hbox{\texttt{rt}}_i$ is less than the expected value $\mu_i$. Thus, the error is the amount by which the expected value differs from  actually observed value. It is standardly assumed that the $\varepsilon_i$ are independently and identically distributed as a normal distribution with mean zero and unknown standard deviation $\sigma_e$. Stan parameterizes the normal distribution by the mean and standard deviation, and we follow that convention here, by writing the distribution of $\varepsilon$ as $\mathrm{N}(0,\sigma _e)$ (the standard notation in statistics is in terms of mean and variance). A consequence of the assumption that the errors are identically distributed is that the distribution of $\varepsilon$ should, at least approximately, have the same shape as the normal distribution. Independence implies that there should be no correlation between the errors---this is not the case in the data, since we have multiple measurements from each subject, and from each item.


\paragraph{Setting up the data}

We now fit the fixed effects Model~\ref{eq:fixef}. 
For the following discussion, please refer to the code in Listings~\ref{fig:Model1code} and \ref{fig:Model1Stancode} in the appendix. First, we read the~Gibson and Wu \cite{gibsonwu} data into a data frame \texttt{rDat} in R, and then subset the critical region (Listing~\ref{fig:Model1code}, lines 2 and 4).
Next, we create a data list \texttt{stanDat} for Stan, which contains the data (Listing 1, line 7). Stan  requires the data to be of type list; this is different from the \texttt{lm} and \texttt{lmer} functions, which assume that the data are in of type data-frame.

\paragraph{Defining the model}

The next step is to write the Stan model in a text file with extension \texttt{.stan}. A Stan model consists of several \textit{blocks}. A block is a set of statements surrounded by brackets and preceded by the block name. We open up a file \texttt{fixEf.stan} in a text editor and write down the first block, the \textit{data block}, which contains the declaration of the variables in the data object \texttt{stanDat} (Listing~\ref{fig:Model1Stancode}, lines 1-5).
The strings \texttt{real} and \texttt{int} specify the data type for each variable. A \texttt{real} variable is a real number, and an \texttt{int} variable is an integer. For instance, \texttt{N} is the integer number of data points. The variables $\hbox{\texttt{so}}$ and $\hbox{\texttt{rt}}$ are arrays of length \texttt{N} whose entries are \texttt{real}. We constrain a variable to take only a subset of the values allowed by its type (e.g. \texttt{int} or \texttt{real}) by specifying in brackets lower and upper bounds (e.g. \texttt{<lower=-1,upper=1>}).

Next, we turn to the \textit{parameters block}, where the parameters are defined (Listing~\ref{fig:Model1Stancode}, lines 6-9).
The fixed effects Model~\ref{eq:fixef} has three parameters: the fixed intercept $\beta _0$, the fixed slope $\beta _1$, and the standard deviation $\sigma _e$ of the error. 
The fixed effects $\beta _0$ and $\beta _1$ are in the vector \texttt{beta} of length two; note that although we called our parameters $\beta_0$ and $\beta_1$ in Model~\ref{eq:fixef}, in Stan, these are contained in a vector with indices 1 and 2, so $\beta_0$ is in \texttt{beta[1]} and $\beta_1$ in \texttt{beta[2]}.
The third parameter, the standard deviation $\sigma _e$ of the error (\texttt{sigma\_e}), is also defined here, and is constrained to have lower bound 0 (Listing~\ref{fig:Model1Stancode}, line 8). Finally,  the \textit{model block} specifies the prior distribution and the likelihood (Listing~\ref{fig:Model1Stancode}, lines 10-15). 

To understand the Stan syntax, compare the Stan code above to the specification of Model~\ref{eq:fixef}. The Stan code literally writes out this model. The block begins with a local variable declaration for \texttt{mu}, which is the mean of $\hbox{\texttt{rt}}$ conditional on whether $\hbox{\texttt{so}}$ is $-1$ for the subject relative condition or $+1$  for the object relative condition. 

The prior distributions on the parameters \texttt{beta} and \texttt{sigma\_e} would ordinarily be declared in the model block. If we don't declare any prior, it is assumed that they have a uniform prior distribution.
Note that the distribution of \texttt{sigma\_e} is truncated at zero because \texttt{sigma\_e} is constrained to be positive (see the declaration \texttt{real<lower=0> sigma\_e;} in the parameters block). So, this means that the error has a uniform prior with lower bound $0$.

In the model block, the for-loop assigns to \texttt{mu} the mean for the log-normal distribution of \texttt{rt[i]}, conditional on the value of the predictor \texttt{so[i]} for relative clause type. The statement \texttt{rt[i] $\sim $ lognormal(mu,sigma\_e)} means that the logarithm of $\hbox{\texttt{rt}}$ is normally distributed with mean \texttt{mu} and standard deviation \texttt{sigma\_e}. One could have equally well log-transformed the reading time and assumed a normal distribution instead of the lognormal. 

\paragraph{Running the model}
We save the file \texttt{fixEf.stan} which we just wrote and fit the model in R with the function \texttt{stan} from the package \texttt{rstan} (Listing~\ref{fig:Model1code}, lines 9 and 10).
This call to the function \texttt{stan} will compile a C++ program which produces samples from the joint posterior distribution of the fixed intercept $\beta _0$, the fixed slope $\beta _1$, and the standard deviation $\sigma _e$ of the error. Here, the function generates four \textit{chains} of samples, each of which contains $2000$ samples of each parameter. Samples $1$ to $1000$ are part of the \textit{warmup}, where the chains settle into the posterior distribution. We analyze samples $1001$ to $2000$. The result is saved to an object \texttt{fixEfFit} of class \texttt{StanFit}.

\begin{figure}
\centering
\begin{knitrout}
\definecolor{shadecolor}{rgb}{0.969, 0.969, 0.969}\color{fgcolor}

{\centering \includegraphics[width=0.75\textwidth]{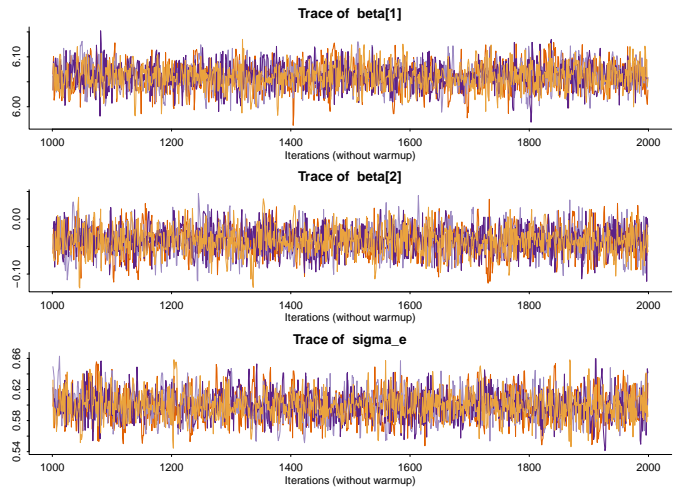} 

}

\end{knitrout}
\caption{Trace plots of the fixed intercept $\beta _0$, the fixed slope $\beta _1$, and the standard deviation $\sigma _e$ of the error for the fixed effects Model~\ref{eq:fixef}.}
\label{fig:traceplot}
\end{figure}

\paragraph{Evaluating model convergence and summarizing results}
The first step after running the above function should be to look at the \textit{trace plot} of each chain after warmup, using the command shown in Listing~\ref{fig:Model1code}, lines 13 and 14.
A trace plot has the chains plotted against the sample ID. In Figure~\ref{fig:traceplot}, we see four different chains plotted against sample number going from $1001$ to $2000$. If the trace plot looks like a ``fat, hairy caterpillar''~\cite{lunn2012bugs} which does not bend, this suggests that the chains have converged to the posterior distribution.

The second diagnostic which we use to assess whether the chains have converged to the posterior distribution is the statistic \texttt{Rhat}. Each parameter has the  \texttt{Rhat} statistic associated with it~\cite{gelman1992inference}; this is essentially the ratio of between-chain variance to within-chain variance (analogous to ANOVA).  The \texttt{Rhat} statistic should be approximately $1\pm 0.1$  if the chain has converged. 
This is shown in the rightmost column of the model summary, see Table~\ref{tab:quantilesGibsonWu}.

Having satisfied ourselves that the chains have converged, we turn to examine this posterior distribution. (If there is an indication that convergence has not happened, then, assuming that the model has no errors in it, increasing the number of samples usually resolves the issue.)

\begin{table}[htdp]
\begin{center}
\begin{tabular}{ccccc}
\hline
parameter        & mean   & 2.5\%  & 97.5\% & $\hat R$\\
\hline
$\hat \beta_0$ & 6.06    & 0.03   &  6.11  &  1\\
$\hat \beta_1$ &  -0.04 & -0.09  &   0.01  &  1\\
$\hat \sigma_e$ &   0.60  & 0.56   &   0.64   &  1\\
\hline
\end{tabular}
\end{center}
\caption{Examining the credible intervals and the R-hat statistic in the Gibson and Wu data.}\label{tab:quantilesGibsonWu}
\end{table}

\begin{figure}
\centering
\begin{knitrout}
\definecolor{shadecolor}{rgb}{0.969, 0.969, 0.969}\color{fgcolor}

{\centering \includegraphics[width=0.75\textwidth]{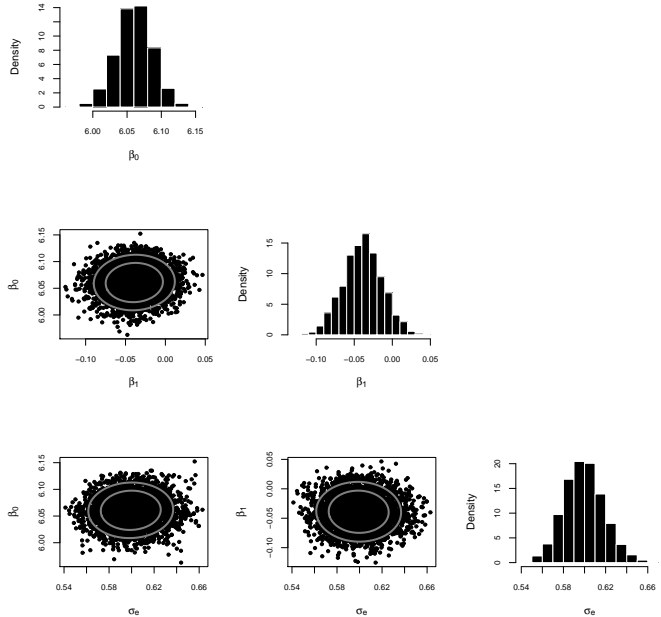} 

}

\end{knitrout}
\caption{Bivariate joint posterior probability distribution of each element of $\theta $ with each other element (lower diagonal) and marginal posterior probability distribution of each element of $\theta $ separately (diagonal). All parameters are on the log scale, but note the difference in length scale between $\beta _1$ on the one hand and $\beta _0$ and $\sigma _e$ on the other.}\label{fig:jointposterior}
\end{figure}

The result of fitting the fixed effects Model~\ref{eq:fixef} is the \textit{joint posterior probability distribution} of the parameters $\beta _0$, $\beta _1$, and $\sigma _e$. The distribution is joint because each of the $(4\text{ chains }\times  1000 \text{ post-warmup iterations }=) 4000$ posterior samples which the call to \texttt{stan} generates is a vector $\theta = ( \beta _0, \beta _1, \sigma _e )^\intercal $ of three model parameters. Thus, the object \texttt{fixEfFit} contains $4000$ parameter vectors $\theta $ which occupy a three dimensional space. Already in three dimensions, the posterior distribution becomes difficult to view in one graph. Figure~\ref{fig:jointposterior} displays the joint posterior probability distribution of the elements of $\theta $ by projecting it down onto planes. In each of the three planes (lower triangular scattergrams) we see how one parameter varies with respect to the other. In the diagonal histograms, we visualize the marginal probability distribution of each parameter separately from the other parameters.

Of immediate interest is the marginal distribution of the slope $\beta _1$. 
Figure~\ref{fig:jointposterior} suggests that the posterior probability density of $\beta _1$ is mainly spread over the interval $(-\infty,0)$. One quantitative way to assess the posterior probability distribution is to examine its quantiles; see Table~\ref{tab:quantilesGibsonWu}. Here, it is useful to define the concept of the \textit{credible interval}. The $(1-\alpha )$\% credible interval contains $(1-\alpha )$\% of the posterior probability density. Unlike the $(1-\alpha )$\% confidence interval from the frequentist setting, the $(1-\alpha )$\% credible interval represents the range within which we are $(1-\alpha )$\% certain that the true value of the parameter lies, given the prior and the data (see \cite{morey2015fallacy} for further discussion on CIs vs credible intervals). A common convention is to use the interval ranging from the $2.5$ to $97.5$ percentiles. We follow this convention and 95\% credible intervals in Table~\ref{tab:quantilesGibsonWu}.

The samples of $\beta_1$ suggests that approximately
94\% of the posterior probability density is below zero, suggesting that there is some evidence that object relatives are easier to process than subject relatives in Chinese, given the Gibson and Wu data. However, since the 95\% credible interval includes 0, we may be reluctant to conclude that object relatives are easier to process.
We will say more about the evaluation of research hypotheses further on. 

\subsection{Varying Intercepts Mixed Effects Model}
\label{subsec:ranint}

The fixed effects Model~\ref{eq:fixef} is inappropriate for the Gibson and Wu data because it does not take into account the fact that we have multiple measurements for each subject and item. As mentioned above, these multiple measurements lead to a violation of the independence of errors assumption. Moreover, the fixed effects coefficients $\beta_0$ and $\beta_1$ represent means over all subjects and items, ignoring the fact that some subjects will be faster and some slower than average; similarly, some items will be read faster than average, and some slower.

In linear mixed models, we take this by-subject and by-item variability into account by adding adjustment terms $u_{0j}$ and $w_{0k}$, which adjust $\beta_0$ for subject $j$ and item $k$. This partially decomposes $\varepsilon _i$ into a sum of the terms $u_{0j}$ and $w_{0k}$, which are adjustments to the intercept $\beta _0$ for the subject $j$ and item $k$ associated with $\hbox{\texttt{rt}}_i$. If subject $j$ is slower than the average of all the subjects, $u_j$ would be some positive number, and if item $k$ is read faster than the average reading time of all the items, then $w_k$ would be some negative number. Each subject $j$ has their own adjustment $u_{0j}$, and each item its own $w_{0k}$. These adjustments $u_{0j}$ and $w_{0k}$ are called \textit{random intercepts} by Pinheiro and Bates \cite{pinheirobates} and
\textit{varying intercepts} by Gelman and Hill \cite{gelmanhill07}, and by adjusting $\beta _0$ by these we account for the variability between speakers, and between items. 

It is standardly assumed that these adjustments are normally distributed around zero with unknown standard deviation: $u_0 \sim \mathrm{N}(0,\sigma _u)$ and $w_0 \sim \mathrm{N}(0,\sigma _w)$; the subject and item adjustments are also assumed to be mutually independent. 
We now have three sources of variance in this model: the standard deviation of the errors $\sigma _e$, the standard deviation of the by-subject random intercepts $\sigma _u $, and the standard deviation of the by-item varying intercepts $\sigma _w $. We will refer to these as variance components.

We now express the logarithm of reading time, which was produced by subjects $j=1,\dots,37$ reading items $k=1,\dots,15$, in conditions $i=1,2$ (1 refers to subject relatives, 2 to object relatives), as the following sum. Notice that we are now using a slightly different way to describe the model, compared to the fixed effects model. We are using indices for subject, item, and condition to identify unique rows. Also, instead of writing $\beta_1 \hbox{\texttt{so}}$, we index $\beta_1$ by the condition $i$. This follows the notation used in the textbook on linear mixed models, written by the authors of \texttt{nlme} \cite{pinheirobates}, the precursor to \texttt{lme4}.

\begin{equation}\label{eq:ranint}
\log \hbox{\texttt{rt}}_{ijk} = \beta _0 + \beta_{1i} + u_{0j} + w_{0k} + \varepsilon_{ijk} 
\end{equation}

Model~\ref{eq:ranint} is an LMM, and more specifically a \textit{varying intercepts model}. The coefficient $\beta_{1i}$ is the one of primary interest; it will have some mean value $-\beta_1$ for subject relatives and $+\beta_1$ for object relatives due to the contrast coding. So, if our posterior mean for $\beta_1$ is negative, this would suggest that object relatives are read faster than subject relatives.

We fit the varying intercepts Model~\ref{eq:ranint} in Stan in much the same way as the fixed effects Model~\ref{eq:fixef}. 
For the following discussion, please consult Listing~\ref{fig:Model2code} for the R code used to run the model, and Listing~\ref{fig:Model2Stancode} for the Stan code.

\paragraph{Setting up the data}

The data which we prepare for passing on to the function \texttt{stan} now includes subject and item information (Listing~\ref{fig:Model2code}, lines 2-8). 
The data block in the Stan code accordingly includes the number \texttt{J}, \texttt{K} of subjects and items, respectively; and the variable \texttt{N} records the number of rows in the data frame.

\paragraph{Defining the model}

Model~\ref{eq:ranint}, shown in Listing~\ref{fig:Model2Stancode}, still has the fixed intercept $\beta _0$, the fixed slope $\beta _1$, and the standard deviation $\sigma _e$ of the error, and we specify these in the same way as we did for the fixed effects Model~\ref{eq:fixef}. In addition, the varying intercepts Model~\ref{eq:ranint} has by-subject varying intercepts $u_{0j}$ for $j=1,\ldots,J$ and by-item varying intercepts $w_{0k}$ for $k=1,\ldots,K$. The standard deviation of $u_0$ is $\sigma _u$ and the standard deviation of $w_0$ is $\sigma _w$. We again constrain the standard deviations to be positive.

The model block places normal distribution priors on the varying intercepts $u_0$ and $w_0$.  We implicitly place uniform priors on \texttt{sigma\_u}, \texttt{sigma\_w}, and \texttt{sigma\_e} by omitting them from the model block. As pointed out earlier for \texttt{sigma\_e}, these prior distributions have lower bound zero because of the constraint \texttt{<lower=0>} in the variable declarations. 

The statement about how each row in the data is generated is shown in Listing~\ref{fig:Model2Stancode}, lines 26-29; here, both the fixed effects and the varying intercepts for subjects and items determine the expected value \texttt{mu}.
The vector \texttt{u} has varying intercepts for subjects. Likewise, the vector \texttt{w} has varying intercepts for items. The for-loop in lines 26-29 now adds \texttt{u[subj[i]] + w[item[i]]} to the mean \texttt{beta[1]} of the distribution of \texttt{rt[i]}. These are subject- and item-specific adjustments to the fixed-effects intercept \texttt{beta[1]}.  The term \texttt{u[subj[i]]} identifies the id of the subject for row $i$ in the data-frame; thus, if $i=1$, then \texttt{subj[1]=1}, and \texttt{item[1]=13} (see Table~\ref{tab:dataframe1}).

\paragraph{Running the model}

We pass the list \texttt{stanDat} of data to \texttt{stan}, which compiles a C++ program to sample from the posterior distribution of Model~\ref{eq:ranint}.
Stan samples from the posterior distribution of the model parameters, including the varying intercepts $u_{0j}$ and $w_{0k}$ for each subject $j=1,\ldots ,J$ and item $k=1,\ldots ,K$. 

It may be helpful to rewrite the model in mathematical form following the Stan syntax (\cite{gelmanhill07} use a similar notation); the Stan statements are slightly different from the way that we expressed Model~\ref{eq:ranint}. Defining $i$ as the row id in the data, i.e., $i=1,\dots, 547$, we can write:

\begin{equation}
\begin{split}
~& \hbox{Likelihood}:\\
~& \mu_i = \beta_0 + u_{[subj[i]]} + w_{[item[i]]} + \beta_1 \times \hbox{\texttt{so}}_i\\
~& \hbox{\texttt{rt}}_i \sim \hbox{LogNormal}(\mu_i,\sigma_e)\\
~& \hbox{Priors}:\\
~& u \sim \hbox{Normal}(0, \sigma_u) \quad w \sim \hbox{Normal}(0,\sigma_w)\\
~& \sigma_e, \sigma_u, \sigma_w \sim \hbox{Uniform}(0,\infty)\\
~& \beta \sim \hbox{Uniform}(-\infty,\infty)\\
\end{split}
\end{equation}

Here, notice that the $i$-th row in the statement for $\mu$ identifies the subject id ranging from $j=1,\dots,37$, and the item id ranging from $k=1,\dots,15$.

\paragraph{Summarizing the results}

The posterior distributions of each of the parameters is summarized in Table~\ref{tab:Model2posterior}. The $\hat R$ values suggest that model has converged. Note also that compared to Model 1, the estimate of $\sigma_e$ is smaller; this is because the other two variance components are now being estimated as well. Note that the 95\% credible interval for the estimate $\hat\beta_1$ includes $0$; thus, there is some evidence that object relatives are easier than subject relatives, but we cannot exclude the possibility that there is no difference in the reading times between the two relative clause types.

\begin{table}[htdp]
\begin{center}
\begin{tabular}{ccccc}
\hline
parameter        & mean   & 2.5\%  &97.5\% & $\hat R$\\
\hline
$\hat \beta_0$ & 6.06 &   5.92  & 6.20  &  1\\
$\hat \beta_1$ &  -0.04    &   -0.08  & 0.01 &  1\\
$\hat \sigma_e$ &   0.52   &     0.49 &  0.55 &  1\\
$\hat \sigma_u$ & 0.25   &    0.19  &  0.34 & 1\\
$\hat \sigma_w$ & 0.20  &       0.12  &  0.32 &     1\\
\hline
\end{tabular}
\end{center}
\caption{The quantiles and the $\hat R$ statistic in the Gibson and Wu data, the varying intercepts model.}\label{tab:Model2posterior}
\end{table}

\subsection{Varying Intercepts, Varying Slopes Mixed Effects Model}\label{subsec:ranslp}

Consider now that subjects who are faster than average (i.e., who have a negative varying intercept) may exhibit greater slowdowns when they read subject relatives compared to object relatives. Similarly, it is in principle possible that items which are read faster (i.e., which have a large negative varying intercept) may show a greater slowdown in subject relatives than object relatives. The opposite situation could also hold: faster subjects may show smaller SR-OR effects, or items read faster may show smaller SR-OR effects.
Although such individual-level variability was not of interest in the original paper by Gibson and Wu, it could be of theoretical interest  (see, for example, \cite{kliegl2010experimental}). Furthermore, as Barr and colleagues \cite{barr2011random} point out, it is in principle desirable to include a fixed effect factor in the random effects as a varying slope if the experiment design is such that subjects see both levels of the factor (cf.\ \cite{BatesEtAlParsimonious}).

In order to express this structure in the LMM, we must make two changes  in the varying intercepts Model~\ref{eq:ranint}. 

\paragraph{Adding varying slopes}

The first change is to let the size of the effect for the predictor \texttt{so} vary by subject and by item. The goal here is to express that some subjects exhibit greater slowdowns in the object relative condition than others. We let effect size vary by subject and by item by including in the model by-subject and by-item varying slopes which adjust the fixed slope $\beta _1$ in the same way that the by-subject and by-item varying intercepts adjust the fixed intercept $\beta _0$. This adjustment of the slope by subject and by item is expressed by adjusting $\beta _1$ by adding two terms $u_{1j}$ and $w_{1k}$. These are \textit{random} or \textit{varying slopes}, and by adding them we account for how the effect of relative clause type varies by subject $j$ and by item $k$. We now express the logarithm of reading time, which was produced by subject $j$ reading item $k$, as the following sum. The subscript $i$ indexes the conditions.

\begin{equation}\label{eq:ranslp}
\log \hbox{\texttt{rt}}_{ijk} = \underbrace{\beta_0 + u_{0j} + w_{0k}}_{\text{varying intercepts}}  + 
\underbrace{\beta_1 + u_{1ij} + w_{1ik}}_{\text{varying slopes}} + \varepsilon_{ijk} 
\end{equation}

\paragraph{Defining a variance-covariance matrix for the random effects}

The second change which we make to Model~\ref{eq:ranint} is to define a covariance relationship between by-subject varying intercepts and slopes, and between by-items intercepts and slopes. This amounts to adding an assumption that the by-subject slopes $u_{1}$ could in principle have some correlation with the by-subject intercepts $u_{0}$; and by-item slopes $w_{1}$ with by-item intercept $w_{0}$. We explain this in detail below.

Let us assume that the adjustments $u_0$ and $u_1$ are normally distributed with mean zero and some variances $\sigma_{u0}^2$ and $\sigma_{u1}^2$, respectively; also assume that $u_0$ and $u_1$ have correlation $\rho_{u}$. It is standard to express this situation by defining a variance-covariance matrix $\Sigma _u$ (sometime this is simply called a variance matrix). 
This matrix has the variances of $u_0$ and $u_1$ respectively along the diagonals, and the covariances on the off-diagonals. (The covariance between two variables X, Y, Cov(X,Y) is defined as the product of their correlation $\rho$ and their standard deviations $\sigma_X$ and $\sigma_Y$: $Cov(X,Y)=\rho\sigma_X \sigma_Y$.)

\begin{equation}\label{eq:covmat}
\Sigma _u
=
\begin{pmatrix}
\sigma _{u0}^2  & \rho _{u}\sigma _{u0}\sigma _{u1}\\
\rho _{u}\sigma _{u0}\sigma _{u1}    & \sigma _{u1}^2\\
\end{pmatrix}
\end{equation}

Similarly, we can define a variance-covariance matrix $\Sigma_w$ for items, using the standard deviations
$\sigma_{w0}$, $\sigma_{w1}$, and the correlation 
$\rho_{w}$.

\begin{equation}\label{eq:covmatw}
\Sigma _w
=
\begin{pmatrix}
\sigma _{w0}^2  & \rho _{w}\sigma _{w0}\sigma _{w1}\\
\rho _{w}\sigma _{w0}\sigma _{w1}    & \sigma _{w1}^2\\
\end{pmatrix}
\end{equation}

The standard way to express this relationship between
the subject intercepts  $u_0$ and slopes $u_1$, and the item intercepts  $w_0$ and slopes $w_1$, is to define a bivariate normal distribution as follows: 

\begin{equation}\label{eq:jointpriordist1}
\begin{pmatrix}
  u_0 \\ 
  u_1 \\
\end{pmatrix}
\sim 
N \left(
\begin{pmatrix}
  0 \\
  0 \\
\end{pmatrix},
\Sigma_{u}
\right)
\quad
\begin{pmatrix}
  w_0 \\ 
  w_1 \\
\end{pmatrix}
\sim 
N\left(
\begin{pmatrix}
  0 \\
  0 \\
\end{pmatrix},
\Sigma_{w}
\right)
\end{equation}

An important point to notice here is that any $n\times n$ variance-covariance matrix has associated with it an $n\times n$ correlation matrix. In the subject variance-covariance matrix $\Sigma_{u}$, the correlation matrix is

\begin{equation}
\begin{pmatrix}
1 & \rho_{01}\\
\rho_{01} & 1\\
\end{pmatrix}
\end{equation}

In a correlation matrix, the diagonal elements will always be $1$, because a variable always has a correlation of $1$ with itself. The off-diagonals will have the correlations between the variables. Note also that, given the variances $\sigma_{u0}^2$ and $\sigma_{u1}^2$, we can always recover the variance-covariance matrix, if we know the correlation matrix. 
This is because of the above-mentioned definition of covariance. 

A correlation matrix can be decomposed into a square root of the matrix, using the Cholesky decomposition. Thus, given a correlation matrix $C$, we can obtain its square root $L$; an obvious consequence is that we can square $L$ to get the correlation matrix $C$ back.  This is easy to illustrate with a simple example. Suppose we have a correlation matrix:

\begin{equation}
C=\begin{pmatrix}
1 & -0.5 \\
-0.5 & 1 \\
\end{pmatrix}
\end{equation}

We can use the Cholesky decomposition function in R, \texttt{chol}, to derive the lower triangular square root $L$ of this matrix. This gives us:

\begin{equation}
L=\begin{pmatrix}
1 & 0 \\
-0.5 & 0.8660254 \\
\end{pmatrix}
\end{equation}

We can confirm that this is a square root by multiplying L with itself to get the correlation matrix back (squaring a matrix is done by multiplying the matrix by its transpose):

\begin{equation}
L L^\intercal=
\begin{pmatrix}
1 & 0 \\
-0.5 & 0.8660254 \\
\end{pmatrix}
\begin{pmatrix}
1 & -0.5 \\
0 & 0.8660254 \\
\end{pmatrix}
=
\begin{pmatrix}
1 & -0.5 \\
-0.5 & 1 \\
\end{pmatrix}
\end{equation}

The reason that we bring up the Cholesky decomposition here is that we will use it to generate the by-subject and by-item adjustments to the intercept and slope fixed-effects parameters. 

\paragraph{Generating correlated random variables using the Cholesky decomposition}

\begin{figure}
\centering
\begin{minipage}{0.45\textwidth}
\begin{knitrout}
\definecolor{shadecolor}{rgb}{0.969, 0.969, 0.969}\color{fgcolor}

{\centering \includegraphics[width=0.75\textwidth]{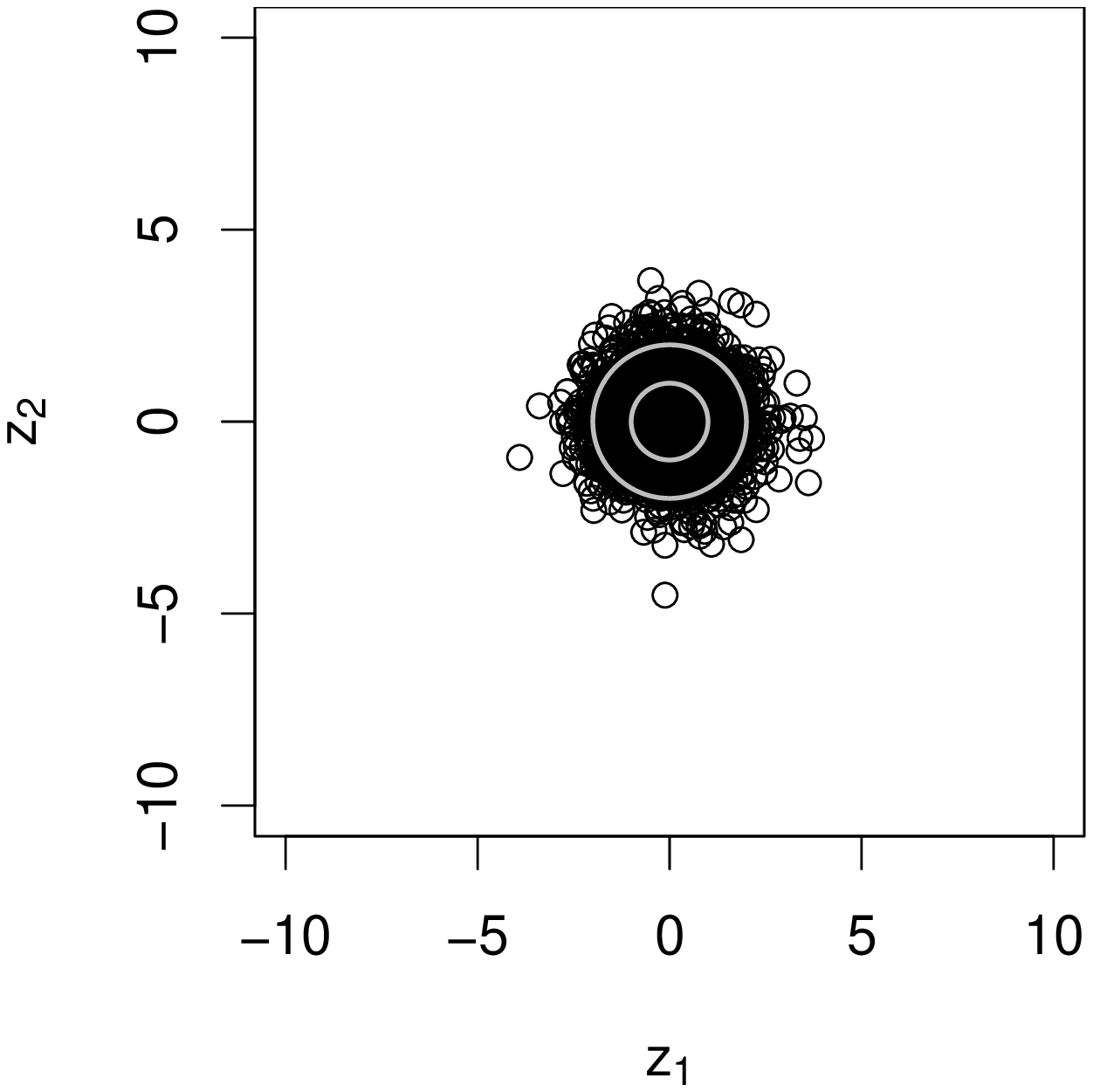} 

}

\end{knitrout}
\end{minipage}
\begin{minipage}{0.45\textwidth}
\begin{knitrout}
\definecolor{shadecolor}{rgb}{0.969, 0.969, 0.969}\color{fgcolor}

{\centering \includegraphics[width=0.75\textwidth]{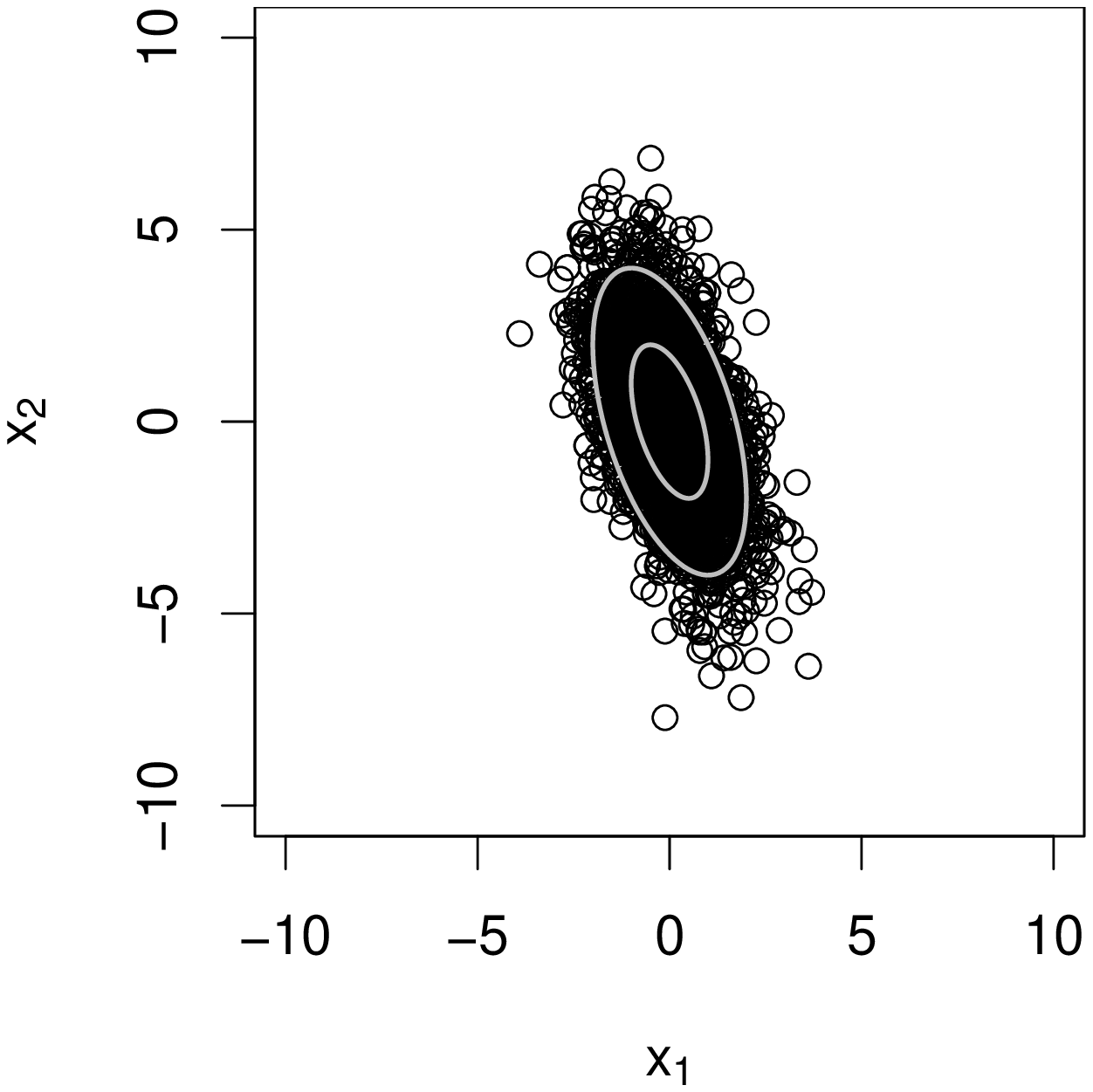} 

}

\end{knitrout}
\end{minipage}
\caption{Uncorrelated random variables $\mathbf{z}=(z_1,z_2)^\intercal $ (left) and correlated random variables $\mathbf{x}=(x_1,x_2)^\intercal $ (right).}\label{fig:xz}
\end{figure}

The by-subject and by-item adjustments are generated using the following standard procedure for generating correlated random variables $\mathbf{x}=(x_1,x_2)$:

\begin{enumerate}
\item
Given a vector of standard deviances (e.g., $\sigma_{u0}, \sigma_{u1}$), create a diagonal matrix:

\begin{equation}
\tau=
\begin{pmatrix}
\sigma_{u0} & 0 \\
0 & \sigma_{u0}\\
\end{pmatrix}
\end{equation}

\item 
Premultiply the diagonalized matrix $\tau$ with the Cholesky decomposition $L$ of the correlation matrix $C$ to get a matrix $\Lambda$.
\item 
Generate values from a random variable $\mathbf{z}=(z_1,z_2)^\intercal $, where $z_1$ and $z_2$ each have independent $\mathrm{N}(0,1)$ distributions (left panel of Figure~\ref{fig:xz}).
\item 
Multiply $\Lambda$ with $\mathbf{z}$; this generates the correlated random variables $\mathbf{x}$ (right panel of Figure~\ref{fig:xz}).
\end{enumerate}

This digression into Cholesky decomposition and the generation of correlated random variables is important to understand for building the Stan model. We will define a vague prior distribution on $L$, and a vague prior on the standard deviances. This allows us to generate the by-subject and by-item adjustments to the fixed effects intercepts and slopes.

\paragraph{Defining the model}

With this background, implementing the varying intercepts, varying slope Model~\ref{eq:ranslp} is straightforward; see Listing~\ref{fig:Model3Stancode} for the code. 
The data block is the same as before.
The parameters block contains several new parameters. 
This time, we have vectors \texttt{sigma\_u} and \texttt{sigma\_w} which are $(\sigma _{u0},\sigma _{u1})^\intercal $ and $(\sigma _{w0},\sigma _{w1})^\intercal $, instead of scalar values as in Model~\ref{eq:ranint}. 
The variables \texttt{L\_u}, \texttt{L\_w}, \texttt{z\_u}, and \texttt{z\_w}, which have been declared in the parameters block, play a role in the \textit{transformed parameters block}, a block which we did not use in the earlier models. The transformed parameters block generates the by-subject and by-item varying intercepts and slopes using the parameters \texttt{sigma\_u}, \texttt{sigma\_w}, \texttt{L\_u}, \texttt{L\_w}, \texttt{z\_u}, and \texttt{z\_w}.  The $J$ pairs of by-subject varying intercepts and slopes are in the rows of the $J\times 2$ matrix \texttt{u}, and the $K$ pairs of by-item varying intercepts and slopes are in the rows of the $K\times 2$ matrix \texttt{w}.

These varying intercepts and slopes are obtained through the statements \texttt{diag\_pre\_multiply(sigma\_u, L\_u) * z\_u} and \texttt{diag\_pre\_multiply(sigma\_w, L\_w) * z\_w}.
This statement generates varying intercepts and slopes from the joint probability distribution of equation~\ref{eq:jointpriordist1}. 
The parameters \texttt{L\_u}, \texttt{L\_w} are the Cholesky decompositions of the subject and item correlation matrices, respectively, and  \texttt{z\_u}, and \texttt{z\_w}  are N(0,1) random variables.

It is helpful to walk through steps 1 to 4 involved in generating the varying intercepts and slopes using the procedure described above for generating correlated random variables. The statement \texttt{diag\_pre\_multiply(sigma\_u,L\_u) * z\_u} computes the transpose matrix product (steps 1 and 2). The right multiplication of this product by \texttt{z\_u}, a matrix of normally distributed random variables (step 3), yields the varying intercepts and slopes (step 4).

\begin{equation}
\begin{split}
\begin{pmatrix}
u_{01} & u_{11} \\
u_{02} & u_{12} \\
\vdots & \vdots \\
u_{0J} & u_{1J}
\end{pmatrix}
&=
\big( \text{diag}(\sigma _{u0}, \sigma _{u1})
\mathbf{L}_u \mathbf{z}_u \big)^\intercal\\
&=
\Bigg(\begin{pmatrix}
\sigma _{u0} & 0\\
0 & \sigma _{01}
\end{pmatrix}
\begin{pmatrix}
\ell _{11} & 0 \\
\ell _{21} & \ell _{22}
\end{pmatrix}
\begin{pmatrix}
z_{11} & z_{12} & \ldots & z_{1J}\\
z_{21} & z_{22} & \ldots & z_{2J}\\
\end{pmatrix}\Bigg)^\intercal
\end{split}
\end{equation} 
 
Turning to the model block, here, we place priors on the parameters declared in the parameters block, and define how these parameters generate $\log \hbox{\texttt{rt}}$ (Listing~\ref{fig:Model3Stancode}, lines 30-42).
The definition of the prior \texttt{L\_u $\sim $ lkj\_corr\_cholesky(2.0)} implicitly places a so-called lkj prior with shape parameter $\eta =2.0$ on the correlation matrices

\begin{equation}
\begin{pmatrix}
1 & \rho _u\\
\rho _u & 1
\end{pmatrix}
\text{ and }
\begin{pmatrix}
1 & \rho _w\\
\rho _w & 1
\end{pmatrix}
\end{equation}

\noindent
where $\rho _u$ is the correlation between the by-subject varying intercept $\sigma _{u0}$ and slope $\sigma _{u1}$ (cf.\ the covariance matrix of Equation~\ref{eq:covmat}) and $\rho _w$ is the correlation between the by-item varying intercept $\sigma _{w0}$ and slope $\sigma _{w1}$. The lkj distribution with shape parameter $\eta =1.0$ is a uniform prior over all $2\times 2$ correlation matrices; it scales up to larger correlation matrices. The parameter $\eta$ has an effect on the shape of the lkj distribution. Our choice of $\eta=2.0$ implies that the correlations in the off-diagonals are near zero, reflecting the prior belief that there is no correlation between intercepts and slopes. 

The statement \texttt{to\_vector(z\_u) $\sim $ normal(0,1)} places a normal distribution with mean zero and standard deviation one on \texttt{z\_u}. The same goes for \texttt{z\_w}. The for-loop assigns to $\hbox{\texttt{mu}}$ the mean of the log-normal distribution from which we draw $\hbox{\texttt{rt[i]}}$, conditional on the value of the predictor \texttt{so[i]} for relative clause type and the subject and item identity.

We can now fit the varying intercepts, varying slopes Model~\ref{eq:ranslp}; see Listing~\ref{fig:preparehusaindata} for the code. We see in the model summary in Table~\ref{tab:Model3posterior} that the model has converged, and that the credible intervals of the parameter of interest, $\beta_1$, still include $0$. In fact, the posterior probability of the parameter being less than $0$ is now $90$\% (this information can be extracted as shown in Listing~\ref{fig:Model3code}, lines 6-8). 

\begin{table}[htdp]
\begin{center}
\begin{tabular}{ccccc}
\hline
parameter        & mean   & 2.5\%  &97.5\% & $\hat R$\\
\hline
$\hat \beta_0$ & 6.06 &   5.92  & 6.21  &  1\\
$\hat \beta_1$ &  -0.04    &   -0.09  & 0.02 &  1\\
$\hat \sigma_e$ &   0.51   &     0.48 &  0.55 &  1\\
$\hat \sigma_{u0}$ & 0.25   &    0.19  &  0.34 & 1\\
$\hat \sigma_{u1}$ & 0.07   &    0.01  &  0.14 & 1\\
$\hat \sigma_{w0}$ & 0.20  &       0.13  &  0.32 &     1\\
$\hat \sigma_{w1}$ & 0.04  &       0.0  &  0.10 &     1\\
\hline
\end{tabular}
\end{center}
\caption{The quantiles and the $\hat R$ statistic in the Gibson and Wu data, the varying intercepts, varying slopes model.}\label{tab:Model3posterior}
\end{table}

\begin{figure}[!htbp]
\centering
\begin{knitrout}
\definecolor{shadecolor}{rgb}{0.969, 0.969, 0.969}\color{fgcolor}

{\centering \includegraphics[width=0.75\textwidth]{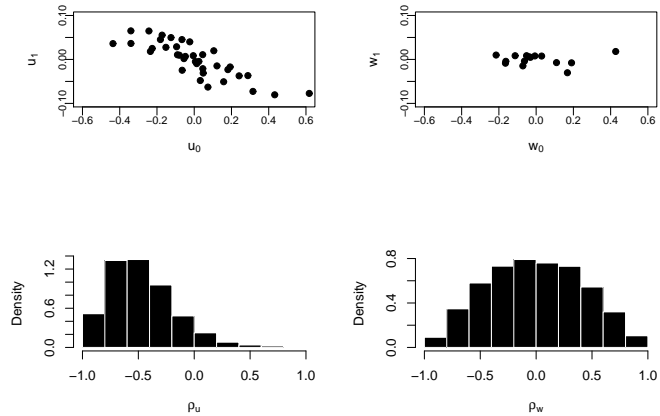} 

}

\end{knitrout}
\caption{The top row shows the relationship between the varying slopes (y-axis) and intercepts (x-axis) for each subject (left panel) and item (right panel). The bottom row shows the posterior distribution of the parameter of correlation between the varying slopes and intercepts for each subject (left panel) and item (right panel).}
\label{fig:lmlist}
\end{figure}

Figure~\ref{fig:lmlist} plots the varying slope's posterior distribution against the varying intercept's posterior distribution for each subject. The correlation between $u_0$ and $u_1$ is negative, as captured by the marginal posterior distributions of the correlation $\rho _u$ between $u_0$ and $u_1$. Thus, Figure~\ref{fig:lmlist} suggests that the slower a subject's reading time is on average, the slower they read object relatives. In contrast, Figure~\ref{fig:lmlist} shows no clear pattern for the by-item varying intercepts and slopes. We briefly discuss inference next.

\section{Inference} \label{sec:bda}
 
Having fit a varying intercepts, varying slopes Model~\ref{eq:ranslp}, we now explain one way to carry out statistical inference, using  credible intervals.  
We have used this approach to draw inferences from data in previously published work (e.g., \cite{FrankEtAl2015}, \cite{HofmeisterVasishth2014}). There are of course other approaches possible for carrying out inference. Bayes Factors are an example; see Lee and Wagenmakers \cite{lee2013bayesian}. Another is to define a Region of Practical Equivalence \cite{kruschke2014doing}. The reader can choose the approach they find the most appealing.

\subsection{Inference using credible intervals} \label{subsec:posteriorintervals}

\begin{figure}
\centering
\begin{knitrout}
\definecolor{shadecolor}{rgb}{0.969, 0.969, 0.969}\color{fgcolor}

{\centering \includegraphics[width=0.75\textwidth]{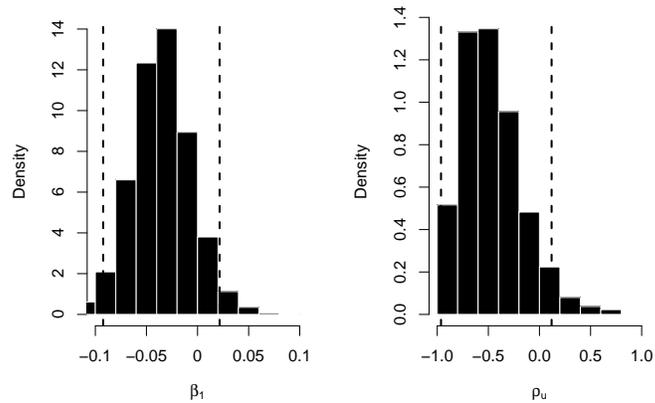} 

}

\end{knitrout}
\caption{Upper and lower bounds on the credible intervals
 (dashed lines) plotted over the marginal posterior distribution of the fixed slope $\beta _1$ (left) and of the correlation $\rho _u$ between the by-subject varying intercepts and varying slopes (right).} \label{fig:hpdinterval}
\end{figure}

The result of fitting the varying intercepts, varying slopes Model~\ref{eq:ranslp} is the posterior distribution of the model parameters. As mentioned above in connection with Models 1-3, direct inference from the posterior distributions is possible. For instance, we can find the posterior probability with which the fixed intercept $\beta _1$ or the correlation $\rho _u$ between by-subject varying intercepts and slopes take on any given value by consulting the marginal posterior distributions whose histograms are shown in Figure~\ref{fig:hpdinterval}. The information conveyed by such graphs can be sharpened by using the $95$\% credible interval, mentioned earlier.  Approximately $95$\% of the posterior density of $\beta _1$ lies between the $2.5$ percentile \ensuremath{-0.092} and the $97.5$ percentile 0.023.
This leads us to conclude that the slope $\beta _1$ for relative clause type $\hbox{\texttt{so}}$ is less than zero with probability $90$\% (see line 8 in Listing~\ref{fig:Model3code}). Since $0$ is included in the credible interval, it is difficult to draw the inference that object relative clauses are read faster than subject relative clauses. However, one could perhaps still make a weak claim to that effect, especially if a lot of evidence has accumulated in other experiments that supports such a conclusion (see \cite{VasishthetalPLoSOne2013} for a more detailed discussion).

What about the correlations between varying intercepts and varying slopes for subject and for item? What can we infer from the analysis about these relationships?
The 95\% credible interval for $\rho _u$ is $($\ensuremath{-1},0.1$)$. Our belief that $\rho _u$ is less than zero is rather uncertain, although we can conclude that $\rho _u$ is less than zero with probability 90\%. There is only weak evidence that subjects who read faster than average exhibit greater slowdowns at the head noun of object relative clauses than subjects who read slower than average. For the by-item varying intercepts and slopes, it is pretty clear that we do not have enough data (15 items) to draw any conclusions. For these data, it probably makes sense to fit a simpler model \cite{BatesEtAlParsimonious}, with only varying intercepts and slopes for subject, and only varying intercepts for items; although there is no harm done here if we fit a model with a full variance-covariance matrix for both subjects and items.

In sum, regarding our main research question, our conclusion here is that we cannot say that object relatives are harder to process than subject relatives, because the credible interval for $\beta_1$ includes $0$. However, one could argue that there is \textit{some} weak evidence in favor of the hypothesis, since the posterior probability of the parameter being negative is approximately $90$\%. 

\section{Example 2: Generalizing the linear mixed model to factorial designs}\label{sec:factorial}

The~Gibson and Wu \cite{gibsonwu} data-set has a two-condition design. This section presents a varying intercepts, varying slopes model for a $2\times 2$ factorial design. Because of the more general matrix formulation we use here, the Stan code can be deployed with minimal changes for much more complex designs, including correlational studies. 

Our example is the $2\times 2$ repeated measures factorial design of~Husain et al \cite{HusainEtAl2014} (Experiment 1), also a self-paced reading study on relative clauses. The dependent variable was the reading time $\hbox{\texttt{rt}}$ of the relative clause verb. The factors were relative clause type, which we code with the predictor $\hbox{\texttt{so}}$ ($\hbox{\texttt{so}}=+1$ for object relatives and $\hbox{\texttt{so}}=-1$ for subject relatives) and distance between the head noun and the relative clause verb, which we code with the predictor $\hbox{\texttt{dist}}$ ($\hbox{\texttt{dist}}=+1$ for far and $\hbox{\texttt{dist}}=-1$ for near). Their interaction is the product of the \texttt{dist} and \texttt{so} contrast vectors, and labeled as the predictor $\hbox{\texttt{int}}$. The $60$ subjects were speakers of Hindi, an Indo-Aryan language spoken primarily in India. The $24$ items were presented in a standard, fully balanced Latin square design. This resulted in a total of $1440$ data points ($60\times 24=1440$). The first few lines from the data frame are shown below.

\begin{table}[htbp]
\centering
\begin{tabular}{rrrrrr}
  \hline
row & subj & item & so & dist & rt \\ 
  \hline
1 &  1 &  14 &  s & n & 1561 \\ 
2 &  1 &  16 &  o & n & 959 \\ 
3 &  1 &  15 &  o & f & 582 \\ 
4 &  1 &  18 &  s & n & 294 \\ 
5 &  1 &   4 &  o & n & 438 \\ 
6 &  1 &  17 &  s & f & 286 \\ 
\vdots  & \vdots & \vdots & \vdots & \vdots & \vdots \\  
1440 &  9 & 13 &  s & f & 516 \\
   \hline
\end{tabular}
\label{tab:dataframe2}
\caption{The first six rows, and the last row, of the data-set of Husain et al.\ (2014, Experiment 1), as they appear in the data frame.}
\end{table}

The theoretical interest is in determining whether relative clause type and distance influence reading time, and whether there is an interaction between these two factors. We use Stan to determine the posterior probability distribution of the fixed effect $\beta _1$ for relative clause type, the fixed effect $\beta _2$ for distance, and their interaction $\beta _3$.

We fit a varying intercepts, varying slopes model to this data-set. This is an extension of Model~\ref{eq:ranslp}.
The grand mean $\beta _0$ of $\log \hbox{\texttt{rt}}$ is adjusted by subject and by item through the varying intercepts $u_0$ and $w_0$, which are unique values for each subject and item respectively. Likewise, the three fixed effects $\beta _1$, $\beta _2$, and $\beta _3$ which are associated with the predictors $\hbox{\texttt{so}}$, $\hbox{\texttt{dist}}$, and $\hbox{\texttt{int}}$, respectively, are adjusted by the by-subject varying slopes $u_1$, $u_2$, and $u_3$ and by-item varying slopes $w_1$, $w_2$, and $w_3$. 

It is more convenient to represent this model in matrix form. We build up the model specification by first noting that, for each subject,  
the by-subject varying intercept $u_0$ and slopes $u_1$, $u_2$, and $u_3$ have a multivariate normal prior distribution with mean zero and covariance matrix $\Sigma _u$. Similarly, for each item, the by-item varying intercept $w_0$ and slopes $w_1$, $w_2$, and $w_3$ have a multivariate normal prior distribution with mean zero and covariance matrix $\Sigma _w$. We can write this as follows:

\begin{equation}
\begin{pmatrix}
  u_0 \\ 
  u_1 \\
  u_2 \\
  u_3
\end{pmatrix}
\sim 
\mathrm{N} \left(
\begin{pmatrix}
  0 \\
  0 \\
  0 \\
  0
\end{pmatrix},
\Sigma_{u}
\right)  
\quad 
\begin{pmatrix}
  w_0 \\ 
  w_1 \\
  w_2 \\
  w_3
\end{pmatrix}
\sim 
\mathrm{N} \left(
\begin{pmatrix}
  0 \\
  0 \\
  0 \\
  0
\end{pmatrix},
\Sigma_{w}
\right)
\end{equation}

The error $\varepsilon $ is assumed to have a normal distribution with mean zero and standard deviation $\sigma _e$. Thus, the varying intercepts, varying slopes here will be the same as Model~\ref{eq:ranslp}, just with two additional predictors $\hbox{\texttt{dist}}$, $\hbox{\texttt{int}}$ along with their associated fixed-effect slopes $\beta _2$, $\beta _3$ and random-effect slopes 
$u_2$, $u_3$, $w_2$, and $w_3$.

We proceed to implement the model in Stan. First we read in the data-set (see Listing~\ref{fig:preparehusaindata}).
Instead of passing the predictors $\hbox{\texttt{so}}$, $\hbox{\texttt{dist}}$, and their interaction $\hbox{\texttt{int}}$ to \texttt{stan} as vectors, as we did with $\hbox{\texttt{so}}$ earlier, we make $\hbox{\texttt{so}}$, $\hbox{\texttt{dist}}$, and $\hbox{\texttt{int}}$ into a design matrix \texttt{X} using the function \texttt{model.matrix} available in R.\footnote{Here, we would like to acknowledge the contribution of Douglas Bates in specifying the model in this general matrix form.}
The first column of the design matrix \texttt{X} consists of all ones. The second column is the predictor $\hbox{\texttt{so}}$ which codes the factor for relative clause type. The third column the predictor $\hbox{\texttt{dist}}$ which codes the factor for distance. The fourth column is the predictor $\hbox{\texttt{int}}$ which codes the interaction between relative clause type and distance.  The model matrix thus consists of a fully factorial $2 \times 2$ design, with blocks of this design repeated for each subject. 
For the full data-set, we could write it very compactly in matrix form as follows:

\begin{equation} \label{eq:factorialmodel}
\mathbf{\log(rt)} = \mathbf{X}\beta + \mathbf{Z}_{u} \mathbf{u} + \mathbf{Z}_{w} \mathbf{w} + \mathbf{\varepsilon} 
\end{equation}

Here,  $\mathbf{X}$ is the $N\times P$ model matrix (with $N=1440$, since we have $1440$ data points; and $P=4$ since we have the intercept plus three other fixed effects), $\mathbf{\beta}$ is a $P\times 1$ vector of fixed effects parameters, $\mathbf{Z}_{u}$ and $\mathbf{Z}_{w}$ are the subject and item model matrices ($N\times P$), and $u$ and $w$ are the by-subject and by-item adjustments to the fixed effects estimates; these are identical to the design matrix $\mathbf{X}$ in the model with varying intercepts and varying slopes included.  For more examples of similar model specifications in Stan, see the R package \texttt{RePsychLing} on github (https://github.com/dmbates/RePsychLing).

Having defined the model, we proceed to assemble the list \texttt{stanDat} of data, relying on the above matrix formulation; please refer to Listing~\ref{fig:preparehusaindata}. The number \texttt{N} of observations, the number \texttt{J} of subjects and \texttt{K} of items, the reading times \texttt{rt}, and the subject and item indicator variables \texttt{subj} and \texttt{item} are familiar from the previous models presented. The integer \texttt{P} is the number of fixed effects (four including the intercept). Model~\ref{eq:factorialmodel} includes a varying intercept $u_{0}$ and varying slopes $u_{1}$, $u_{2}$, $u_{3}$ for each subject, and so the number \texttt{n\_u} of by-subject random effects equals \texttt{P}. Likewise, Model~\ref{eq:factorialmodel} includes a varying intercept $w_{0}$ and varying slopes $w_{1}$, $w_{2}$, $w_{3}$ for each item, and so the number \texttt{n\_w} of by-item random effects also equals \texttt{P}. 
The data block contains the corresponding variables. We declare the fixed effects design matrix \texttt{X} as an array of \texttt{N} row vectors whose components are the predictors associated with the \texttt{N} reading times. Likewise for the subject and item random effects design matrices \texttt{Z\_u} and \texttt{Z\_w}, which correspond to $\mathbf{Z}_{u}$ and $\mathbf{Z}_{w}$ respectively in Model~\ref{eq:factorialmodel}. 
The vector \texttt{beta} contains the fixed effects $\beta _0$, $\beta _1$, $\beta _2$, and $\beta _3$. The matrices \texttt{L\_u}, \texttt{L\_w} and the arrays \texttt{z\_u}, \texttt{z\_w} of vectors (not to be confused with the design matrices \texttt{Z\_u} and \texttt{Z\_w}) will generate the varying intercepts and slopes $u_0$, \dots , $u_3$ and $w_0$, \dots , $w_3$, using the procedure described for Model~\ref{eq:ranslp}. The vector \texttt{sigma\_u} contains the standard deviations of the by-subject varying intercepts and slopes $u_0$, \dots , $u_3$, and the vector \texttt{sigma\_w} contains the standard deviations of the by-item varying intercepts and slopes $w_0$, \ldots , $w_3$. The variable \texttt{sigma\_e} is the standard deviation $\sigma _e$ of the error $\varepsilon$.
The transformed parameters block generates the by-subject intercepts and slopes $u_0$, \dots , $u_3$ and the by-item intercepts and slopes $w_0$, \dots, $w_3$.

We place lkj priors on the random effects correlation matrices through the \texttt{lkj\_corr\_cholesky(2.0)} priors on their Cholesky factors \texttt{L\_u} and \texttt{L\_w}. We implicitly place uniform priors on the fixed effects $\beta _0$, \dots , $\beta _3$, the random effects standard deviations $\sigma _{u0}$, \dots , $\sigma _{u3}$, and $\sigma _{w0}$, \dots, $\sigma _{w3}$ and the error standard deviation $\sigma _e$ by omitting any prior specifications for them in the model block. We specify the likelihood with the probability statement that \texttt{rt[i]} is distributed log-normally with mean \texttt{X[i] * beta + Z\_u[i] * u[subj[i]] + Z\_w[i] * w[item[i]]} and standard deviation \texttt{sigma\_e}.
The next step towards model-fitting is to pass the list \texttt{stanDat} to \texttt{stan}, which compiles a C++ program to sample from the posterior distribution of the model parameters.

\begin{figure}
\centering
\begin{knitrout}
\definecolor{shadecolor}{rgb}{0.969, 0.969, 0.969}\color{fgcolor}

{\centering \includegraphics[width=0.75\textwidth]{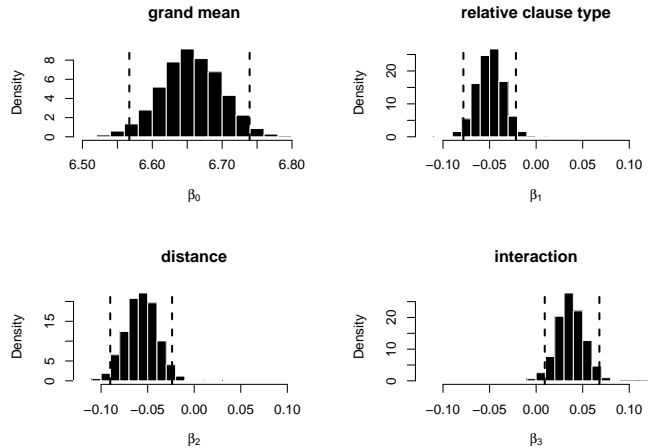} 

}

\end{knitrout}
\caption{Marginal posterior distribution and HPD intervals of the fixed effects grand mean $\beta _0$, slope $\beta _1$ for relative clause type, slope $\beta _2$ for distance, and interaction $\beta _3$. All fixed effects are on the log-scale.}\label{fig:factorialfixefposterior}
\end{figure}

Figure~\ref{fig:factorialfixefposterior} plots histograms of the marginal posterior distribution of the fixed effects. The HPD interval of the fixed effect $\hat\beta _1$ for relative clause type is entirely below zero. This is evidence that object relatives are read faster than subject relatives. The HPD interval of the fixed effect $\hat\beta _2$ for distance is also entirely below zero. This is evidence of a slowdown when the verb (where reading time was measured) is closer to the head noun of the relative clause. The HPD of the interaction $\hat\beta _3$ between relative clause type and distance is greater than zero, which is evidence for a greater slowdown on subject relatives when the distance between the verb and head noun is short.

A major advantage of the above matrix formulation is that we do not need to write a new Stan model for a future repeated measures factorial design. All we have to do now is define the design matrix $X$ appropriately, and include it (along with appropriately defined $Z_u$ and $Z_w$ for the subjects and items random effects) as part of the data specification that is passed to Stan. 

\section{Concluding remarks, and further reading}

We hope that this tutorial has given the reader a flavor of what it would be like to fit Bayesian linear mixed models. There is of course much more to say on the topic, and we hope that the interested reader will take a look at some of the excellent books that have recently come out. We suggest below a sequence of reading that we found helpful.
A good first general textbook is by Gelman and Hill \cite{gelmanhill07}; it begins with the frequentist approach and only later transitions to Bayesian models. The forthcoming book by Mcelreath \cite{rethinking} is also excellent. 
For those looking for a psychology-specific introduction, the books by Kruschke \cite{kruschke2014doing} and Lee and Wagenmakers \cite{lee2013bayesian} are to be recommended, although for the latter the going might be easier if the reader has already looked at Gelman and Hill \cite{gelmanhill07}.
As a second book, \cite{lunn2012bugs} is recommended; it provides many interesting and useful examples using the BUGS language, which are discussed in exceptionally clear language. Many of these books use the BUGS syntax~\cite{lunn2000winbugs}, which the probabilistic programming language JAGS \cite{plummer2011jags} also adopts; however, Stan code for these books is slowly becoming available on the Stan home page (https://github.com/stan-dev/example-models/wiki). 
For those with introductory calculus, a slightly more technical introduction to Bayesian methods by Lynch~\cite{lynch2007introduction} is an excellent choice. Finally, the textbook by Gelman and colleagues~\cite{Gelman14} is the definitive modern guide, and provides a more advanced treatment. 

\section*{Acknowledgements}

We are grateful to the developers of Stan (in particular, Andrew Gelman, Bob Carpenter) and members of the Stan mailing list for their advice regarding model specification. Douglas Bates and Reinhold Kliegl have helped considerably over the years in improving our understanding of LMMs from a frequentist perspective. We also thank Edward Gibson for releasing his published data. Titus von der Malsburg and Bruno Nicenboim provided useful comments on a draft.

\bibliographystyle{plain}
\bibliography{SorensenVasishth}

\appendix

\singlespacing

\begin{listing}
\begin{Verbatim}[numbers=left,frame=single,fontfamily=courier,fontsize=\footnotesize]
## read in data:
rDat <- read.table ( "gibsonwu2012data.txt" , header=TRUE )
## subset critical region:
rDat <- subset( rDat , region == "headnoun" )

## create data as list for Stan, and fit model:
stanDat <- list( rt = rDat$rt, so = rDat$type, N = nrow(rDat) )
library(stan)
fixEfFit <- stan ( file = "fixEf.stan" , data = stanDat , 
               iter = 2000 , chains = 4 )

## plot traceplot, excluding warm-up:
traceplot( fixEfFit , pars = c("beta","sigma_e"),
          inc_warmup = FALSE)

## examine quantiles of posterior distributions:
print( fixEfFit , pars = c("beta","sigma_e") , 
       probs = c(0.025,0.5,0.975))

## examine quantiles of parameter of interest:
beta1 <- extract ( fixEfFit , pars=c("beta[2]"))$beta
print ( signif ( quantile ( beta1,probs = c(0.025,0.5,0.975)) 
        , 2))
\end{Verbatim}
\caption{Code for the fixed effects Model 1.}\label{fig:Model1code}
\end{listing}

\newpage

\begin{listing}
\begin{Verbatim}[numbers=left,frame=single,fontfamily=courier,fontsize=\footnotesize]
data {
  int<lower=1> N;                //number of data points
  real rt[N];                    //reading time
  real<lower=-1,upper=1> so[N];  //predictor
}
parameters {
  vector[2] beta;            //intercept and slope
  real<lower=0> sigma_e;     //error sd
}
model {
  real mu;
  for (i in 1:N){                   // likelihood
    mu <- beta[1] + beta[2]*so[i];
    rt[i] ~ lognormal(mu,sigma_e);
  }
}
\end{Verbatim}
\caption{Stan code for the fixed effects Model 1.}\label{fig:Model1Stancode}
\end{listing}

\newpage

\begin{listing}
\begin{Verbatim}[numbers=left,frame=single,fontfamily=courier,fontsize=\footnotesize]
## format data for Stan:
stanDat<-list(subj=as.integer(factor(rDat$subj)),
              item=as.integer(factor(rDat$item)),
              rt=rDat$rt,
              so=rDat$so,
              N=nrow(rDat),
              J=length(unique(rDat$subj)),
              K=length(unique(rDat$item)))

## Sample from posterior distribution:
ranIntFit <- stan(file="ranInt.stan", data=stanDat, 
                  iter=2000, chains=4)
## Summarize results:
print(ranIntFit,pars=c("beta","sigma_e","sigma_u","sigma_w"),
      probs=c(0.025,0.5,0.975))

beta1 <- extract(ranIntFit,pars=c("beta[2]"))$beta
print(signif(quantile(beta1,probs=c(0.025,0.5,0.975)),2))

## Posterior probability of beta1 being less than 0:
mean(beta1<0)                  
\end{Verbatim}
\caption{Code for running Model 2, the varying intercepts model.}\label{fig:Model2code}
\end{listing}

\newpage

\begin{listing}
\begin{Verbatim}[numbers=left,frame=single,fontfamily=courier,fontsize=\footnotesize]
data {
  int<lower=1> N;                  //number of data points
  real rt[N];                      //reading time
  real<lower=-1,upper=1> so[N];    //predictor
  int<lower=1> J;                  //number of subjects
  int<lower=1> K;                  //number of items
  int<lower=1, upper=J> subj[N];   //subject id
  int<lower=1, upper=K> item[N];   //item id
}

parameters {
  vector[2] beta;            //fixed intercept and slope
  vector[J] u;               //subject intercepts
  vector[K] w;               //item intercepts
  real<lower=0> sigma_e;     //error sd
  real<lower=0> sigma_u;     //subj sd
  real<lower=0> sigma_w;     //item sd
}

model {
  real mu;
  //priors
  u ~ normal(0,sigma_u);    //subj random effects
  w ~ normal(0,sigma_w);    //item random effects
  // likelihood
  for (i in 1:N){
    mu <- beta[1] + u[subj[i]] + w[item[i]] + beta[2]*so[i];
    rt[i] ~ lognormal(mu,sigma_e);
  }
}
\end{Verbatim}
\caption{Stan code for running Model 2, the varying intercepts model.}\label{fig:Model2Stancode}
\end{listing}

\newpage

\begin{listing}
\begin{Verbatim}[numbers=left,frame=single,fontfamily=courier,fontsize=\footnotesize]
ranIntSlpFit <- stan(file="ranIntSlp.stan", data = stanDat, 
                     iter=2000, chains = 4)

## posterior probability of beta 1 being less
## than 0:
beta1 <- extract(ranIntSlpFit,pars=c("beta[2]"))$beta
print(signif(quantile(beta1,probs=c(0.025,0.5,0.975)),2))
mean(beta1<0)
\end{Verbatim}
\caption{Code for running Model 3, the varying intercepts, varying slopes model.}\label{fig:Model3code}
\end{listing}

\newpage

\begin{listing}
\begin{Verbatim}[numbers=left,frame=single,fontfamily=courier,fontsize=\footnotesize]
data {
  int<lower=1> N;                  //number of data points
  real rt[N];                      //reading time
  real<lower=-1,upper=1> so[N];    //predictor
  int<lower=1> J;                  //number of subjects
  int<lower=1> K;                  //number of items
  int<lower=1, upper=J> subj[N];   //subject id
  int<lower=1, upper=K> item[N];   //item id
}

parameters {
  vector[2] beta;                  //intercept and slope
  real<lower=0> sigma_e;           //error sd
  vector<lower=0>[2] sigma_u;      //subj sd
  vector<lower=0>[2] sigma_w;      //item sd
  cholesky_factor_corr[2] L_u;
  cholesky_factor_corr[2] L_w;
  matrix[2,J] z_u;
  matrix[2,K] z_w;
}

transformed parameters{
  matrix[2,J] u;
  matrix[2,K] w;
  
  u <- diag_pre_multiply(sigma_u,L_u) * z_u;  //subj random effects
  w <- diag_pre_multiply(sigma_w,L_w) * z_w;  //item random effects
}

model {
  real mu;
  //priors
  L_u ~ lkj_corr_cholesky(2.0);
  L_w ~ lkj_corr_cholesky(2.0);
  to_vector(z_u) ~ normal(0,1);
  to_vector(z_w) ~ normal(0,1);
  //likelihood
  for (i in 1:N){
    mu <- beta[1] + u[1,subj[i]] + w[1,item[i]] 
          + (beta[2] + u[2,subj[i]] + w[2,item[i]])*so[i];
    rt[i] ~ lognormal(mu,sigma_e);
  }
}
\end{Verbatim}
\caption{The Stan code for Model 3, the varying intercepts, varying slopes model.}\label{fig:Model3Stancode}
\end{listing}

\newpage

\begin{listing}
\begin{Verbatim}[numbers=left,frame=single,fontfamily=courier,fontsize=\footnotesize]
rDat<-read.table("HusainEtAlexpt1data.txt",header=TRUE)
rDat$subj <- with(rDat,factor(subj))
rDat$item <- with(rDat,factor(item))

X <- unname(model.matrix(~1+so+dist+int, rDat))

stanDat <- within(list(),
{
  N<-nrow(X)
  P <- n_u <- n_w <- ncol(X)
  X <- X
  Z_u <- X 
  Z_w <- X
  J <- length(levels(rDat$subj))
  K <- length(levels(rDat$item))
  rt <- rDat$rt
  subj <- as.integer(rDat$subj)
  item <- as.integer(rDat$item)
}
)
factorialFit <- stan(file="factorialModel.stan",
                     data=stanDat,
                     iter=2000, chains=4)                     
\end{Verbatim}
\caption{Preparation of data for analyzing the Husain et al.\ data-set, and running the model.}\label{fig:preparehusaindata}
\end{listing}

\newpage

\begin{listing}
\begin{Verbatim}[numbers=left,frame=single,fontfamily=courier,fontsize=\footnotesize]
data {
  int<lower=0> N;               //no trials
  int<lower=1> P;               //no fixefs
  int<lower=0> J;               //no subjects
  int<lower=1> n_u;             //no subj ranefs
  int<lower=0> K;               //no items
  int<lower=1> n_w;             //no item ranefs
  int<lower=1,upper=J> subj[N]; //subject indicator
  int<lower=1,upper=K> item[N]; //item indicator
  row_vector[P] X[N];           //fixef design matrix
  row_vector[n_u] Z_u[N];       //subj ranef design matrix
  row_vector[n_w] Z_w[N];       //item ranef design matrix
  vector[N] rt;                 //reading time
}
parameters {
  vector[P] beta;               //fixef coefs
  cholesky_factor_corr[n_u] L_u;  //cholesky factor of subj ranef corr matrix
  cholesky_factor_corr[n_w] L_w;  //cholesky factor of item ranef corr matrix
  vector<lower=0>[n_u] sigma_u; //subj ranef std
  vector<lower=0>[n_w] sigma_w; //item ranef std
  real<lower=0> sigma_e;        //residual std
  vector[n_u] z_u[J];           //subj ranef
  vector[n_w] z_w[K];           //item ranef
}
transformed parameters {
  vector[n_u] u[J];             //subj ranefs
  vector[n_w] w[K];             //item ranefs
  {
    matrix[n_u,n_u] Sigma_u;    //subj ranef cov matrix
    matrix[n_w,n_w] Sigma_w;    //item ranef cov matrix
    Sigma_u <- diag_pre_multiply(sigma_u,L_u);
    Sigma_w <- diag_pre_multiply(sigma_w,L_w);
    for(j in 1:J)
      u[j] <- Sigma_u * z_u[j];
    for(k in 1:K)
      w[k] <- Sigma_w * z_w[k];
  }
}
model {
  //priors
  L_u ~ lkj_corr_cholesky(2.0);
  L_w ~ lkj_corr_cholesky(2.0);
  for (j in 1:J)
    z_u[j] ~ normal(0,1);
  for (k in 1:K)
    z_w[k] ~ normal(0,1);
  //likelihood
  for (i in 1:N)
    rt[i] ~ lognormal(X[i] * beta + 
                      Z_u[i] * u[subj[i]] + 
                      Z_w[i] * w[item[i]], 
                      sigma_e);
}
\end{Verbatim}
\caption{Stan code for Husain et al data.}\label{fig:Stancodehusaindata}
\end{listing}

\end{document}